\documentclass{pasj00}
\draft

\begin{document}
\SetRunningHead{S. Katsuda et al.}{The Ejecta Structure of the Cygnus Loop}  
\Received{2007/05/31}%{yyyy/mm/dd}
\Accepted{2007/08/08}%{yyyy/mm/dd}

\title{Asymmetric Ejecta Distribution of the Cygnus Loop revealed
  with Suzaku}

%%% begin:list of authors
% Do NOT capitalize all letters in "textsc".
%\author{Satoru \textsc{Katsuda}, Hiroshi \textsc{Tsunemi}}
%,Emi \textsc{Miyata}, \author{Koji \textsc{Mori} Masaaki \textsc{Namiki}, Norbert \textsc{Nemes},
%Naohisa \textsc{Anabuki}, and Hiroyuki \textsc{Uchida}} % 
%  \thanks{Example: Present Address is xxxxxxxxxx}}
%\affil{Department of Earth and Space Science, Graduate School of
%  Science, Osaka University, 1-1 Machikaneyama, Toyonaka, Osaka
%  560-0043, Japan
%  }\email{katsuda@ess.sci.osaka-u.ac.jp, tsunemi@ess.sci.osaka-u.ac.jp}
%  miyata@ess.sci.osaka-u.ac.jp, namiki@ess.sci.osaka-u.ac.jp,
%  nnemes@ess.sci.osaka-u.ac.jp, anabuki@ess.sci.osaka-u.ac.jp,
%  uchida@ess.sci.osaka-u.ac.jp} 

%\author{Eric \textsc{Miller}}
%\affil{Kavli Institute for Astrophysics and Space Research, Massachusetts
%Institute of Technology, Cambridge, MA 02139,
%U.S.A.}\email{milleric@space.mit.edu}

%
%\affil{$Department of Applied Physics, Faculty of Engineering,
%University of Miyazaki, 889-2192, Japan}\email{mori@astro.miyazaki-u.ac.jp}

%%% end:list of authors

%%% Please use the following style in case that sorting by 
%%% affilation is impossible. 
%
 \author{
   Satoru \textsc{Katsuda}\altaffilmark{1},
   Hiroshi \textsc{Tsunemi}\altaffilmark{1},
   Emi \textsc{Miyata}\altaffilmark{1},
   Koji \textsc{Mori}\altaffilmark{2},
   Masaaki \textsc{Namiki}\altaffilmark{1},
   Norbert \textsc{Nemes}\altaffilmark{1}
\and
   Eric D. \textsc{Miller}\altaffilmark{3}
}
 \altaffiltext{1}{Department of Earth and Space Science, Graduate School of
  Science, Osaka University, 1-1 Machikaneyama, Toyonaka, Osaka
  560-0043}
 \email{katsuda@ess.sci.osaka-u.ac.jp}
 \altaffiltext{2}{Department of Applied Physics, Faculty of Engineering,
University of Miyazaki, 889-2192}
 \altaffiltext{3}{Kavli Institute for Astrophysics and Space Research,
   Massachusetts Institute of Technology, Cambridge, MA 02139, U.S.A.}

%% `\KeyWords{}' always has to be placed before `\maketitle'.
 %Do NOT move this preamble from here!

\maketitle

\begin{abstract}
We observed a linearly sliced area of the Cygnus Loop from the
north-east to the south-west with Suzaku in seven pointings.  After
dividing the entire fields 
of view (FOV) into 119 cells, we extracted spectra from all of the cells and
performed spectral analysis for them. We then applied both
one- and two-component non-equilibrium ionization (NEI)
models for all of the spectra, finding that almost all were
significantly better fitted by the two-component NEI
model rather than the one-component NEI model.
Judging from the abundances, the high-$kT_\mathrm{e}$ component must be
the ejecta component, while the low-$kT_\mathrm{e}$ component comes from the
swept-up matter.  Therefore, the ejecta turn out to be distributed
inside a large area (at least our FOV) of the Cygnus Loop.
We divided the entire FOV into northern and southern parts, and 
found that the ejecta distributions were asymmetric to the geometric
center: the ejecta of Si, S, and Fe seem to be distributed more in
the south than in the north of the Cygnus Loop by a factor of
$\sim$2. The degree of ejecta-asymmetry is consistent with that
expected by recent supernova explosion models.
\end{abstract}

\section{Introduction}

The asymmetry of a supernova (SN) explosion is considered to be a key
point to understand the mechanism driving an SN explosion.
The spectropolarimetry of young SNe revealed a non-spherical core of the SN
explosion.  The degree of departure from spherical 
symmetry is considered to be anti-correlated with the mass of the hydrogen
envelope, i.e., the remaining hydrogen shell of the progenitor star
(e.g., Leonard et al.\ 2006).  
In the extreme case of departure from spherical symmetry, for
example Type-IIn SN 1998S, the major-to-minor axis 
ratio of the explosion core is expected to be larger than 2.5,
suggesting a highly asymmetric explosion mechanism, possibly mediated by
jets (Wang et al.\ 2001).   
From theoretical aspects, recent SN explosion models generally show
aspherical ejecta distributions may also be produced by convection (e.g.,
Herant et al.\ 1994; Burrows et al.\ 1995; Janka \& Muller 1996) as well as
hydrodynamical instabilities (e.g., Kifonidis et al.\ 2006;
Burrows et al.\ 2007; Janka et al.\ 2007) during SN explosions.

After a SN explosion, the ejecta initially expand without deceleration
(so-called ``free expansion'') behind the forward shock wave, which
sweeps up the interstellar medium (ISM).  When the mass of the swept-up
matter becomes comparable to that of the ejecta, the expansion is
decelerated by the swept-up matter.  The deceleration initiates a
reverse shock propagating into the ejecta.  Both the swept-up matter and
the ejecta are hot enough to emit X-rays.  The swept-up matter forms a
shell structure, while the ejecta fill its interior.  Therefore, we can
attempt to address the ejecta distributions by observing the structure
of supernova 
remnants (SNRs) in X-rays.  In fact, ejecta-dominated X-ray emission are
detected in many SNRs (e.g., Cas A: Hughes et al.\ 2000; Tycho:
Decourchelle et al.\ 2001; Vela: Tsunemi et al.\ 1999; G292.0+1.8:
Park et al.\ 2001). Recent deep Chandra X-ray observations of Cas A SNR  
have revealed a bipolar structure of Si-rich ejecta (Hwang et al.\ 2004;
Laming et al.\ 2003), suggesting an asymmetric explosion with bipolar jets.

The Cygnus Loop is a nearby (540\,pc: Blair et al.\ 2005) proto-typical
middle-aged ($\sim$10000 yrs) SNR.  A number of pieces of evidence 
that the SN explosion of the Cygnus Loop occurred in a preexisting cavity 
(e.g., McCray 1979; Levenson et al.\ 1999), suggesting that the Cygnus
Loop was produced by a core-collapse SN.  Since it is an evolved SNR,
the ejecta-material is embedded in swept-up matter.  Miyata et al.\ (1994)
observed the northeast (NE) shell of the Loop with ASCA, and revealed the
metal deficiency there.  Recent Suzaku observations
of this region confirm the metal deficiency there (Miyata et al.\ 2007). 
Although the reason for the metal deficiency is not yet
understood, it can be safely stated that contamination of the
ejecta-material is negligible in the rim regions.  In contrast to the
rim regions, ASCA detected evidence of Si-, S-, and Fe-rich plasma at
the center portion of the Cygnus Loop (Miyata et al.\ 1998), which is
thought to be ejecta.  The relative abundances of the ejecta support
that the idea that the Cygnus Loop was the result from a core-collapse
SN, and that the progenitor mass is estimated to be 25\,M$_\odot$
based on the central portion of the Cygnus Loop (Miyata et al.\ 1998).  
Recent XMM-Newton observations across the Cygnus Loop from the NE
rim to the SW rim revealed ejecta distributed inside of
$\sim0.85\,R_s$ of the Cygnus Loop, where $R_s$ is the shock radius
(Tsunemi et al.\ 2007).   The relative abundances inferred for the total
ejecta are almost consistent with those expected for the core-collapse
SN, whose progenitor mass is 13\,M$_\odot$ (Katsuda \& Tsunemi 2008), which
can be considered to be more reliable than that estimated from the ASCA
observation, since the coverage of the total FOV of the XMM-Newton
observations is about an order of magnitude larger than that of the ASCA
observation, giving a more representative sample of the ejecta
abundances to compare with theoretical models.

Since the Cygnus Loop is very large in apparent size
(\timeform{2.5D}$\times$\timeform{3.5D}: Levenson et al.\ 1997;
Aschenbach \& Leahy 1999), it is an ideal target to study the spatial
distribution of the ejecta material.  In order to reveal ejecta
structures in this SNR, we observed the Cygnus Loop from the NE rim to
the SW rim with Suzaku (Mitsuda et al.\ 2007).  We selected our fields of
view (FOV) so as to cover just southern regions of the XMM-Newton
observation path since the distribution of Si-, S-, and Fe-rich ejecta is
suggested to be elongated toward the south of the Loop (Tsunemi et
al.\ 2007). 

\begin{figure*}
  \begin{center}
    \FigureFile(80mm,80mm){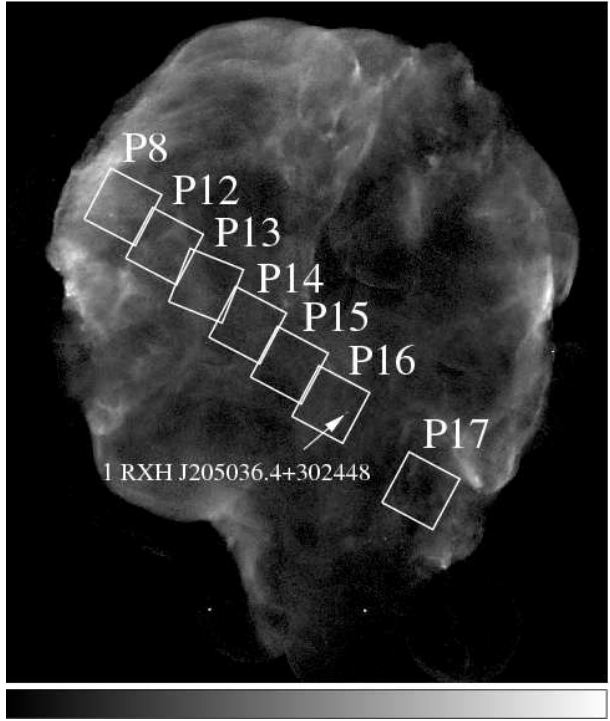}
    %%% \FigureFile(width,height){filename}
  \end{center}
  \caption{ROSAT HRI image of the entire Cygnus Loop. The Suzaku FOV
  (P8, P12, P13, P14, P15, P16, and P17) are shown as white
 rectangles. }\label{fig:hri_image} 
\end{figure*}

%\newpage

\section{Observations and Data Screening}

\begin{figure*}
  \begin{center}
    \FigureFile(160mm,160mm){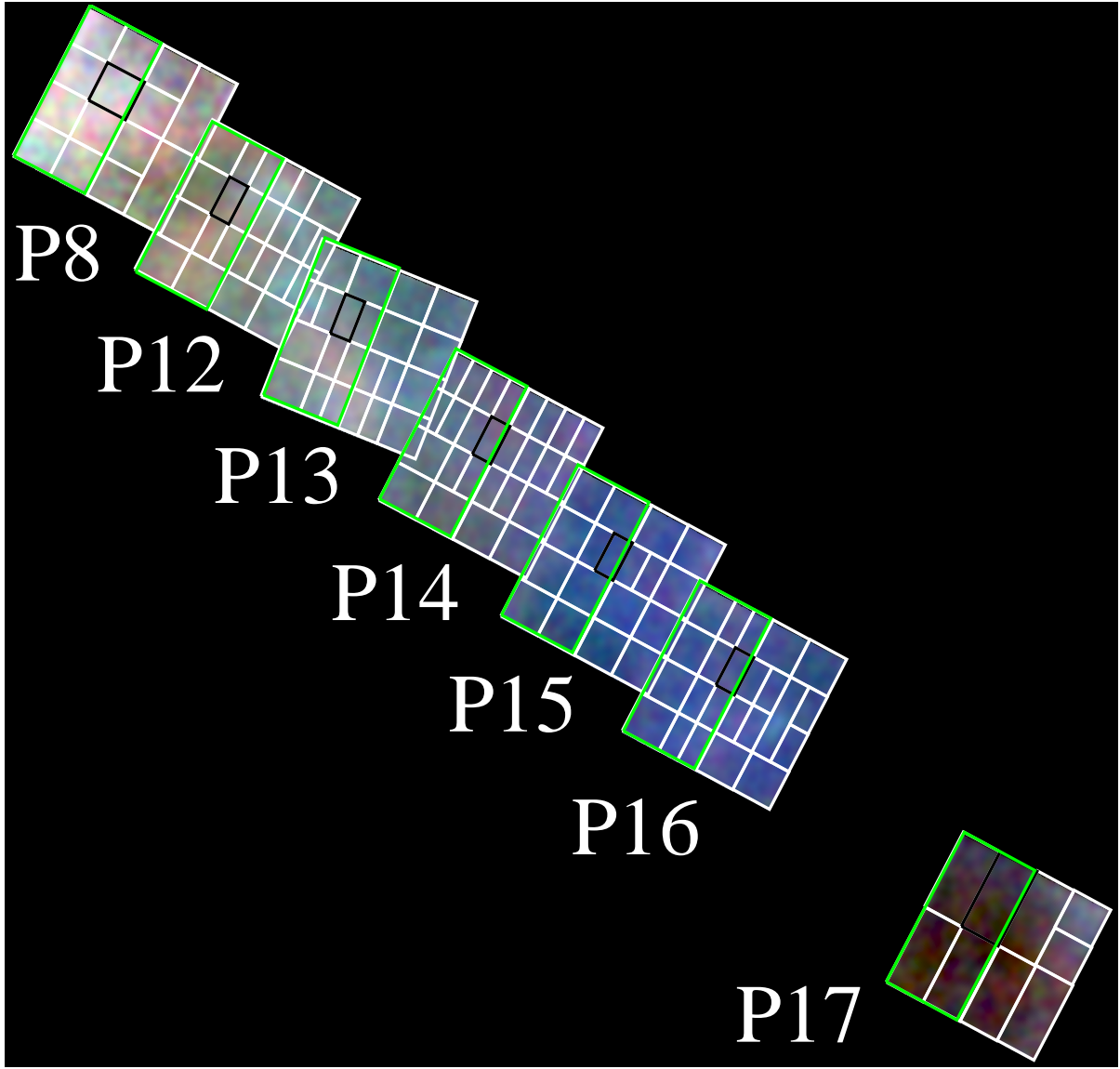}
    %%% \FigureFile(width,height){filename}
  \end{center}
  \caption{Three-color image merged by seven XIS FOV
 (Red: O {\scshape VII} K$\alpha$, Green: Ne {\scshape IX} K$\alpha$,
 Blue: Fe L ).  The data were binned by 8 pixels and smoothed by
 a Gaussian kernel of $\sigma = 25^{\prime\prime}$.  The effects
 of exposure, vignetting, and contamination are corrected.  Small
 rectangles were the cells where we extracted spectra.  We show
 example spectra from black cells for $<$2\,keV and green cells for
 $>$2\,keV in figure~\ref{fig:ex_spec}.}  
	\label{fig:xis1_image} 
\end{figure*}

The observations comprised seven pointings: P8, P12, P13,
P14, P15, P16, and P17.  The FOV of our Suzaku observations are
shown in figure~\ref{fig:hri_image}.  We employed revision 1.2 of the   
cleaned event data. We excluded data taken in the low cut-off
rigidity $<$\,6\,GV. As a background, we used a spectrum obtained
from the Lockman Hole because its observation date, 2006 May 17 was
close to those of the Cygnus Loop.  
Obs. IDs, the nominal point, roll angle, observation date, and
the effective exposure times after the screening
are summarized in table~\ref{obs}. 
Figure~\ref{fig:xis1_image} shows a merged XIS1 (back-illuminated CCD; BI
CCD) three-color image.  Red, green, and blue colors correspond to
narrow energy bands of 
0.54--0.59\,keV (O {\scshape VII} K$\alpha$), 0.88--0.94\,keV (Ne
{\scshape IX} K$\alpha$), and 0.69--0.85\,keV (Fe L), respectively.
We can see strong blue color around the center portion (i.e., P15 and
P16), whereas red and green colors are enhanced in the NE regions
(i.e., P8 and P12).  For spectrum fitting, we used photons
in the energy ranges of 0.2--3.0\,keV and 0.4--3.0\,keV for XIS1 and
XIS0, 2, 3 (front-illuminated CCD; FI CCD), respectively.

\begin{table*}
 \begin{center}
 \caption{Information of observations of the Cygnus Loop and Lockman Hole.}
  \begin{tabular}{lcccc}
\hline
Obs. ID &Coordinate (RA, DEC)&Roll &Obs. Date& Effective Exposure\\
\hline
\multicolumn{5}{c}{Cygnus Loop}\\
501028010 (P8) & 314.005, 31.464 & \timeform{58.5D} & 2006.05.13 & 4.7\,ks\\
501029010 (P12) & 313.751, 31.263 & \timeform{58.7} & 2006.5.09 & 13.2\,ks\\
501030010 (P13) & 313.498, 31.061 & \timeform{58.9} & 2006.05.10 & 13.9\,ks\\
501031010 (P14) & 313.245, 30.859 & \timeform{59.1} & 2006.5.12 & 18.2\,ks\\
501032010 (P15) & 312.993, 30.653& \timeform{59.3} & 2006.05.25 & 17.4\,ks\\
501033010 (P16) & 312.745, 30.450 & \timeform{59.5} & 2006.05.22 & 20.0\,ks\\
501034010 (P17) & 312.207, 30.005 & \timeform{60.0} & 2006.05.22 & 13.9\,ks\\
\hline
\multicolumn{5}{c}{Lockman Hole}\\
101002010 & 162.938, 57.267 & \timeform{32.7D} & 2006.05.17 & 69.0\,ks\\
\hline
\label{obs}
  \end{tabular}
 \end{center}
\end{table*}

\section{Spatially Resolved Spectral Analysis}

We divided the entire FOV into 119 rectangular small cells, indicated in
figure~\ref{fig:xis1_image}, such that each cell contained 2500--5000
photons for XIS0 to equalize the statistics.  The sizes of small cells
ranged from 1/32 FOV to 1/8 FOV. We then extracted spectra
from them.  Generally, count rates above 2\,keV were so low that the
statistics were too poor in each cell.  It is quite difficult to constrain
the abundance of S.  Therefore, by dividing each FOV into the NE part and
the SW part, we accumulated photons in the energy range
above 2\,keV from the NE or SW half, where each small cell was included,
whereas we extracted spectra below 2\,keV from each small cell.  
Example regions in each FOV, a black cell for $<$2\,keV and a green cell for
$>$2\,keV, are shown in figure~\ref{fig:xis1_image}. 
In this way, we obtained much better constraints on the S 
abundance for each cell than those determined by fitting
entire-energy-band spectra extracted from each small cell.  
The S abundances were different for each small cell due to the spectral
differences below 2\,keV. To generate the response matrix file (RMF) and the 
ancillary response file (ARF), we employed {\tt xisrmfgen} and {\tt
xissimarfgen} (Ishisaki et al.\ 2007) (version 2006-10-26),
respectively. The low-energy efficiency of the XIS shows degradation
caused by contaminants accumulated on the optical blocking filter
(Koyama et al.\ 2007), which was taken into
account generating of the ARF file.  

\begin{figure*}
  \begin{center}
    \FigureFile(150mm,200mm){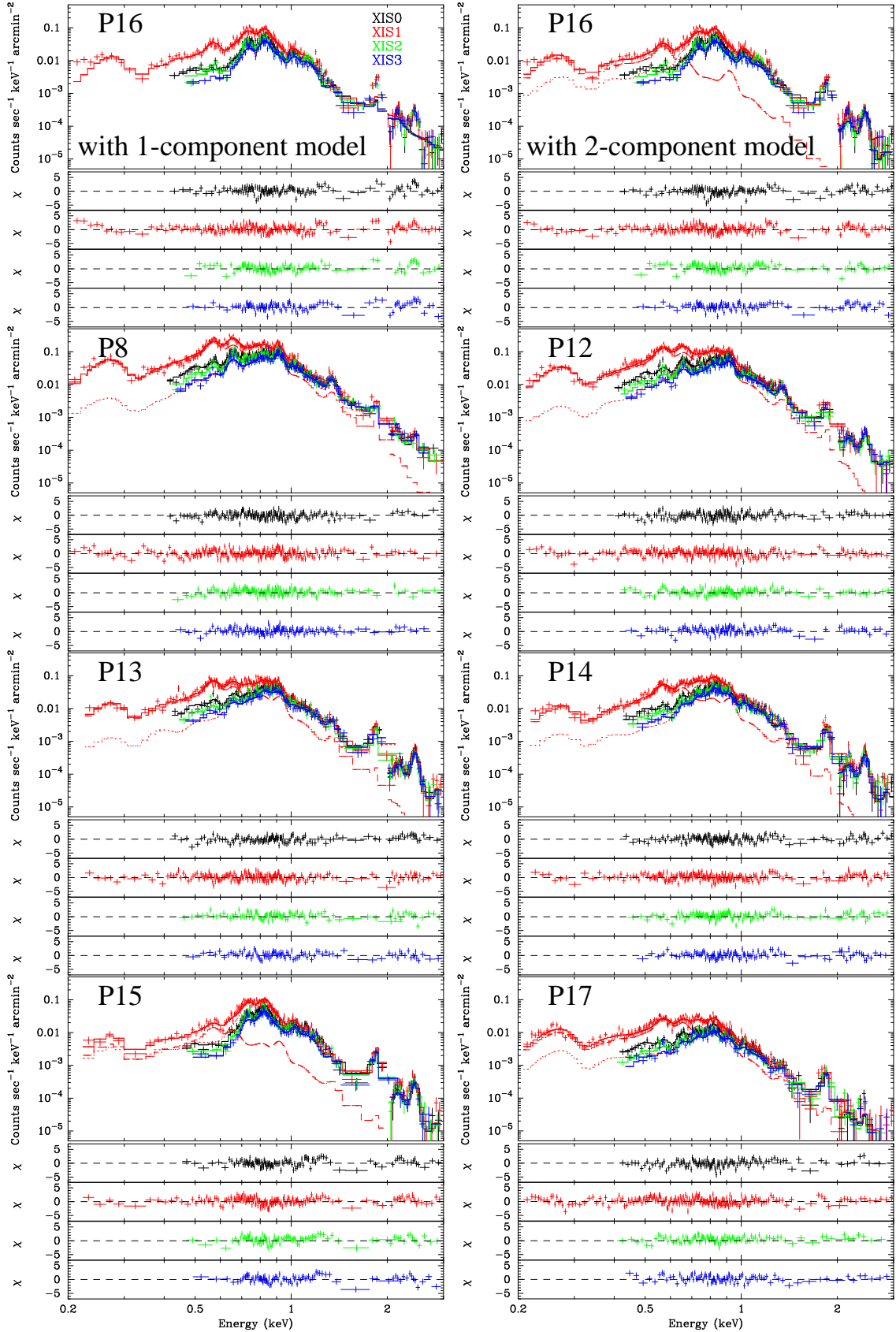}
    %%% \FigureFile(width,height){filename}
  \end{center}
  \caption{Top-left panel: X-ray spectra extracted from the black cell in P16
 indicated in figure~\ref{fig:xis1_image}.  The best-fit curves for
 one-component NEI model are shown with solid lines and the lower panels
 show the residuals. The small gap around 2\,keV is due to the fact that
 we extracted spectra below 2\,keV from each small cell, while above
 2\,keV from half of each FOV.
 Top-right panel: Same as top left but with two-component
 NEI best-fit models.  The contribution of each component is
 shown by the dotted lines only for XIS1.  Other panels: Same as
 top-right panel but from black cells in P8, P12, P13, P14, P15, and P17, 
 respectively.}\label{fig:ex_spec}   
\end{figure*}

Firstly, all spectra were fitted by an absorbed non-equilibrium
ionization (NEI) model with a single component [the wabs;
Morrison \& McCammon 1983 and the VNEI model (NEI version 2.0);
e.g., Borkowski et al.\ 2001 in XSPEC v\,11.3.1]. 
Free parameters are the hydrogen column density in cm$^{-2}$,
$N_\mathrm{H}$; electron temperature in keV, $kT_\mathrm{e}$; the
ionization time, $\tau$, where $\tau$ is the electron density in
cm$^{-3}$ times the elapsed time in sec after the shock heating; the
emission measure, EM (EM$=\int n_\mathrm{e}n_\mathrm{H} dl$, where
$n_\mathrm{e}$ and $n_\mathrm{H}$ are the number densities of
electrons and hydrogens in cm$^{-3}$, respectively and $dl$ is the
plasma depth in cm); the abundances of C, N, O, Ne, Mg, Si, S, Fe, 
and Ni. We set abundance of Ni to be equal to that of Fe.  
The abundances of other elements included in the VNEI model (i.e., Ar
and Ca) were fixed to the solar values (Anders \& Grevesse 1989).
An example spectrum from the black cell in P16, shown in
Figure~\ref{fig:xis1_image}, is presented in figure~\ref{fig:ex_spec},
indicated as P16 (left) together with the best-fit model.  
This model gave us quite good fits, except for the energy bands around
Si- and S-K lines where we can see a large discrepancy between our data
and the model.  Since the emission from both 
the ejecta and the swept-up matter were detected as a projection effect
around the center portion of the Cygnus Loop (Miyata et al.\ 1998;
Tsunemi et al.\ 2007; Katsuda \& Tsunemi 2008), it is natural to
consider that we need at least two (i.e., the swept-up matter and the
ejecta) components to reproduce our data.  
%Also, recent XMM-Newton observations across the Cygnus Loop from NE rim
%to SW rim showed that spectra in outer regions of $R > 0.7\,R_s$ were
%represented by one component NEI model with low abundances, which was hence
%considered to be originated from the swept-up ISM, on the other hand
%most of the spectra in $R <0.7\,R_s$ required additional component with
%high abundances whose origin is naturally considered to be the ejecta.

We then applied a two-component NEI model for all spectra. In this
model, $kT_\mathrm{e}$, $\tau$, and EM are free parameters for both
components.  $N_\mathrm{H}$ is also left as a free parameter, but
tied in the two components.  Assuming that the swept-up matter surrounds
the ejecta, we fixed the metal abundances for the swept-up matter
component to those of the NE rim regions of the Cygnus Loop, where we
expect no ejecta component (Uchida et al.\ 2006).  The metal abundances
are described in the footnote of table \ref{tab:ex_param}.  Then, we
left the metal abundances of O(=C=N), Ne, Mg, Si, S, and Fe(=Ni) for
the ejecta component as free parameters.  We confined the values of
$N_\mathrm{H}$ in the range from 0.01 to 0.06
$\times10^{22}$\,cm$^{-2}$ (Inoue et al.\ 1980; Miyata et al.\ 2007).  
Figure~\ref{fig:ex_spec} shows example spectra from seven black cells
shown in figure~\ref{fig:xis1_image}.  We summarize the best-fit
parameters for the example spectra in table~\ref{tab:ex_param}. 
Based on table~\ref{tab:ex_param}, we found that the two components 
clearly showed different temperature with each other.  The
low-$kT_\mathrm{e}$ component was responsible for the swept-up matter while
the high-$kT_\mathrm{e}$ component was responsible for another component.
We found that the metal abundances for at least one element of the
high-$kT_\mathrm{e}$ component were about an order-of-magnitude higher
than those for the low-$kT_\mathrm{e}$ component, which clearly 
showed that the high-$kT_\mathrm{e}$ component represented the ejecta
component.  In this way, we performed spectral fittings for all 
spectra from 119 small cells.  Applying the $F$-test with a significance
level of 99\%, we found that this model gave us better fits than those
for one-component NEI models for almost all of the spectra [e.g., from
figure~\ref{fig:ex_spec} P16 (left) to P16 (right)].  We obtained fairly
good fits for all spectra by the two-component NEI model (reduced
$\chi^2 < 1.7$). 
If we consider the calibration uncertainty of the energy scale
($\pm$5\,eV; Koyama et al.\ 2007), the values of reduced $\chi^2$ become
lower than 1.4.  Nonetheless, the fits are not acceptable for many
spectra from a statistical point of view, which suggests that 
our model is too simple.  However, we believe that the two-component
model is a good approximation to represent our data.  Taking into
consideration that the spectra from almost all cells require the
two-component model, the ejecta turned out to be 
distributed inside a large fraction (at least our FOV) of the Cygnus Loop.  
Since the outer edge of our FOV is located around $0.85\,R_s$,
our result matches the view from XMM-Newton observations that
the ejecta are distributed inside $\sim0.85\,R_s$ of the Cygnus Loop.

\begin{table*}
%\tabletypesize{\tiny}
  \begin{center}
  \caption{Spectral-fit parameters for the example
   spectra.}\label{tab:ex_param} 
    \begin{tabular}{lccccccc}
\hline
Parameter & P8 & P12 & P13 & P14 & P15 & P16 & P17\\
\hline
$N_\mathrm{H}$[$\times10^{22}$cm$^{-2}$] \dotfill & 0.044$^{+0.005}_{-0.003}$ &
     0.049$^{+0.005}_{-0.003}$ & 0.060$\pm$0.004&0.060$\pm$0.004& 
     0.049$^{+0.006}_{-0.003}$& 0.060$\pm$0.003&
     0.031$\pm$0.002\\ 
\hline
 \multicolumn{8}{c} {High temperature component}\\
$kT_\mathrm{e1}$[keV] \dotfill & 0.55$\pm$0.01 & 0.73$^{+0.03}_{-0.01}$
     & 0.60$\pm$0.04 & 0.60$\pm$0.04 & 0.38$\pm$0.01 &
     0.39$\pm$0.01& 0.35$\pm$0.01 \\
O(=C=N) \dotfill& 0.9$\pm$0.1 & 0.6$\pm$0.1 & 0.47$^{+0.09}_{-0.08}$ &
     1.2$\pm$0.4 &
     0.30$\pm$0.05 & 0.5$\pm$0.1 & 0.20$\pm$0.05\\
Ne \dotfill& 0.8$\pm$0.1& 0.7$\pm$0.1 & 0.8$\pm$0.2 & 0.7$\pm$0.2&
     0.29$\pm$0.06 & 0.46$^{+0.07}_{-0.08}$ & 0.15$\pm$0.05\\
Mg \dotfill& 0.5$\pm$0.1& 0.5$\pm$0.1 & 0.8$\pm$0.2 & 0.6$\pm$0.2&
     0.16$\pm$0.09 & 0.14$\pm$0.09 & 0.14$\pm$0.09 \\ 
Si \dotfill& 0.6$\pm$0.2& 1.0$\pm$0.2 & 5.1$\pm$0.6 & 6.1$\pm$0.7&
     3.6$\pm$0.4 &  4.1$\pm$0.4 & 2.7$^{+0.5}_{-0.6}$\\ 
S \dotfill& 1.2$\pm$0.4& 2.8$\pm$0.5 & 10$\pm$1 & 11$\pm$1&
     7.1$\pm$0.9 &  7.5$\pm$0.8 & 4$\pm$3\\
Fe(=Ni) \dotfill&0.62$^{+0.06}_{-0.08}$& 1.1$\pm$0.1 & 3.3$\pm$0.1
     &3.9$\pm$0.2& 
     2.04$\pm$0.04 & 2.30$\pm$0.05 & 0.75$\pm$0.03\\ 
log$(\tau_1 /\mathrm{cm}^{-3}\,\mathrm{s})$\dotfill
     &11.2$^{+0.04}_{-0.09}$& 10.73$^{+0.06}_{-0.04}$ & 11.11$^{+0.07}_{-0.11}$
     &11.5$\pm$0.1 &
     11.76$^{+0.10}_{-0.07}$ & 11.76$^{+0.08}_{-0.09}$ & 12.0$^{+1.7}_{-0.2}$\\ 
EM$^\dagger_1$[$\times10^{19}$ cm$^{-5}$]\dotfill& 0.22$\pm$0.01 &
     0.072$^{+0.006}_{-0.005}$ & 0.043$\pm$0.001 &0.061$\pm$0.002 &
     0.166$\pm$0.003 & 
     0.131$\pm$0.002 & 0.102$\pm$0.003 \\ 
\hline
 \multicolumn{8}{c} {Low temperature component}\\
$kT_\mathrm{e2}$[keV] \dotfill & 0.24$\pm$0.01 & 0.23$\pm$0.01
     & 0.23$\pm$0.01 & 0.21$\pm$0.01& 0.18$\pm$0.01 & 0.15$\pm$0.01&
     0.22$\pm$0.01 \\ 
Abundances \dotfill&  \multicolumn{7}{c} {(fixed to those
     determined for the NE rim of the Cygnus Loop)$^\ddagger$}\\
log$(\tau_2 /\mathrm{cm}^{-3}\,\mathrm{s})$\dotfill
     &11.33$\pm$0.04 & 11.3$\pm$0.1 & 11.21$\pm$0.05 &
     11.67$^{+0.09}_{-0.08}$& 
     13.1$^{+0.6}_{-1.1}$ & 12.0$^{+1.7}_{-0.2}$ & 11.58$\pm$0.04\\ 
EM$^\dagger_2$[$\times10^{19}$ cm$^{-5}$]\dotfill& 4.4$^{+0.2}_{-0.1}$ &
     2.8$^{+0.1}_{-0.2}$ & 1.91$\pm$0.06 & 2.16$^{+0.10}_{-0.09}$ &
     0.67$\pm$0.05 & 1.5$^{+0.2}_{-0.1}$& 0.63$\pm$0.01 \\  
\hline
$\chi^2$/d.o.f. \dotfill & 772/643 & 749/617 & 524/492 & 653/576 & 654/459 & 751/525 & 769/523\\
      \hline
\\[-8pt]
  \multicolumn{3}{@{}l@{}}{\hbox to 0pt{\parbox{140mm}{\footnotesize
     \par\noindent 
\footnotemark[$*$]Other elements are fixed to those of solar values.\\
     The values of abundances are multiples of solar value.\\  The errors
     are in the range $\Delta\,\chi^2\,<\,2.7$ on one parameter.   
\par\noindent 
\footnotemark[$^\dagger$]EM denotes the emission measure, $\int
     n_\mathrm{e} n_\mathrm{H} dl$. 
\par\noindent 
\footnotemark[$^\ddagger$]C$=$0.27, N$=$0.10, O$=$0.11, Ne$=$0.21,
     Mg$=$0.17, Si$=$0.34, S$=$0.17, and Fe($=$Ni)$=$0.20
     (Uchida et al.\ 2006).\\ 
\par\noindent 
}\hss}}
    \end{tabular}
  \end{center}
\end{table*}

\begin{figure*}
  \begin{center}
    \FigureFile(160mm,100mm){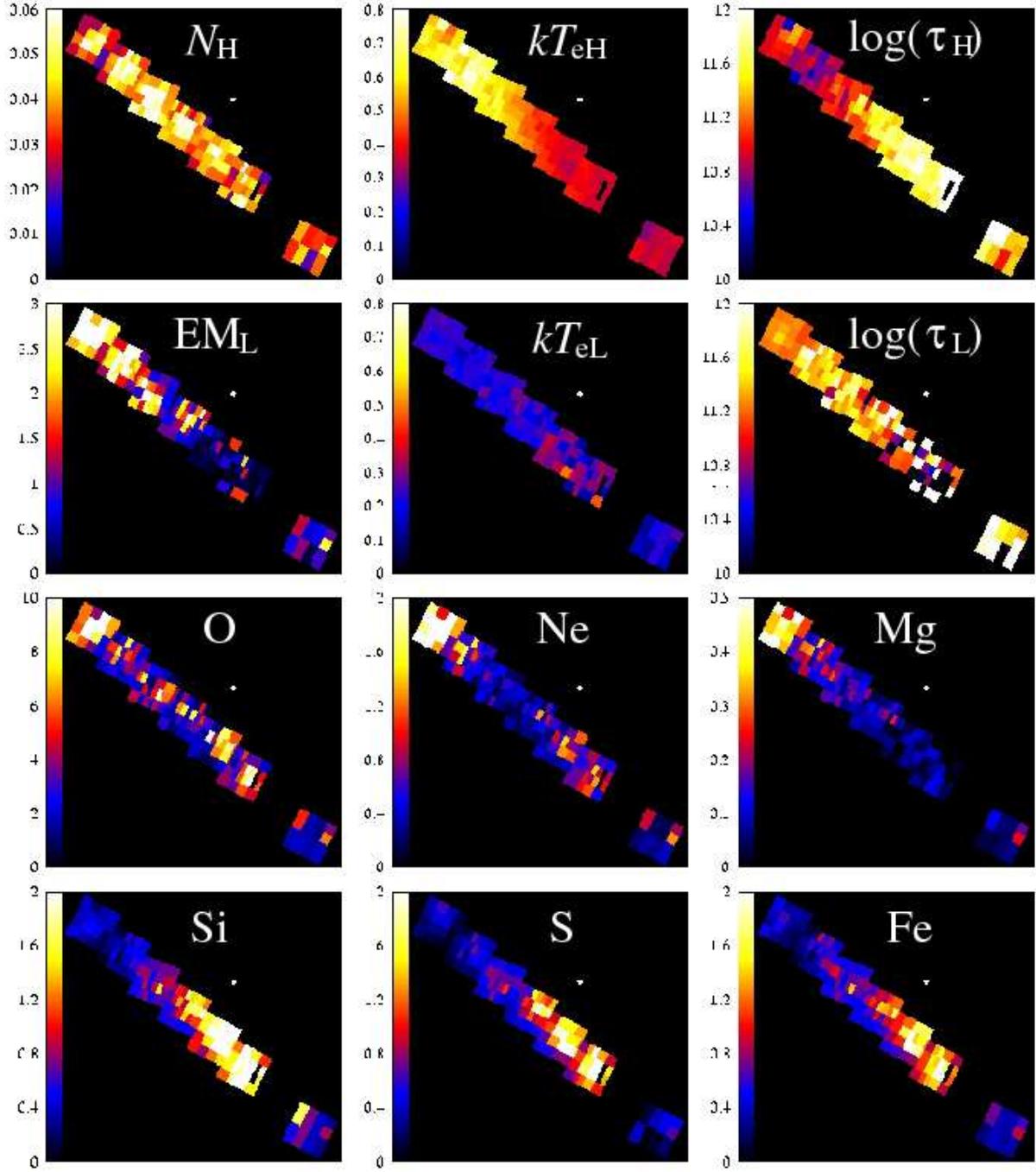}
    %%% \FigureFile(width,height){filename}
  \end{center}
  \caption{Maps of the best-fit parameters.  The values of
 $N_\mathrm{H}$, $kT_\mathrm{e}$, EM$_\mathrm{L}$ are in units of
 $10^{22} \mathrm{cm}^{-2}$, keV, and $10^{19} \mathrm{cm}^{-5}$,
 respectively.  Lower six maps show EMs of O, Ne, Mg, Si, S, and Fe for
 the high-$kT_\mathrm{e}$ component in units of $10^{14}
 \mathrm{cm}^{-5}$.  We adjusted the color code, such that we can see the
 differences in each figure. }\label{fig:param}
\end{figure*}

Figure~\ref{fig:param} shows maps of the best-fit parameters
obtained from the two-component NEI model.  The subscript, H, denotes
the high-$kT_\mathrm{e}$, i.e., the ejecta component while L denotes the
low-$kT_\mathrm{e}$, i.e., the swept-up matter component.   
We showed EMs of O, Ne, Mg, Si, S, and Fe for the ejecta component.
We marked the geometric center of the Cygnus Loop 
(RA=\timeform{20h51m19s}, DEC=$31^\circ02^\prime48^{\prime\prime}$ [J2000]:
Ku et al.\ 1984) as a white dot in each figure.  
There is a point source, 1RXH J$205036.4+302448$, within P16 (see,
figure~\ref{fig:hri_image}).  We excluded the responsible cell from
our analysis, which can be seen as a black box in the map 
of log($\tau_\mathrm{H}$) shown in figure~\ref{fig:param}. $N_\mathrm{H}$s
tend to show relatively large values around the center in each FOV to
those in the edge regions. 
Since such $N_\mathrm{H}$ variations must be due to the contamination
problem of the XIS (Koyama et al.\ 2007), the
current model of a spatial variation of the contaminants accumulated on
the optical blocking filter should be modified properly.  We found that
the value of $kT_\mathrm{e}$ for the swept-up matter component is
$\sim$0.2\,keV in our entire FOV other than P15 and P16, where we can see
slightly higher values ($\sim$0.3\,keV). The values of $kT_\mathrm{e}$
for the ejecta component show a significant variation from the SW
($\sim$0.35\,keV) to the NE ($\sim$0.7\,keV) portion in our FOV as
Miyata et al.\ (2000) already noted the hard X-ray-emitting region in
the NE portion by ASCA 
GIS observations and suggested that the hard X-ray might come from the
ejecta. The collisional ionization equilibrium (CIE) has already been
established for both components in most cells in P17, which seems
consistent with the results from Chandra observations of the SW rim of
the Cygnus Loop (Leahy 2004).

 We noticed a clear unti-correlation between $kT_\mathrm{eH}$ and
$\tau_\mathrm{H}$, and a correlation between $\tau_\mathrm{H}$ and
EMs of Si, S, and Fe in the ejecta component.  We examined the
systematic effect between the ionization age (and/or temperature) and the
distributions of the Si-, S-, and Fe-ejecta.  We re-fitted the example
spectrum in P15, employing  the same two-component model used in this
paper, but with a fixed ionization age of
$1\times10^{11}\mathrm{cm}^{-3}\,\mathrm{s}$ for the ejecta
component (which is the typical value in P8, 12, 13).  We noticed a
slight increase of $kT_\mathrm{eH}$ that is still significantly lower
than that in the northern part of our FOV.  We found
that the emission measures of Si, S, and Fe were, respectively,
reduced by 30\%, 30\%, and 15\% from those 
obtained in the case that we left the ionization age as a free parameter.
These values are still higher than those obtained in P8, 12, and 13 by
$\sim$50\%.  Therefore, we can safely conclude that the EMs of
Si-, S-, and Fe-ejecta in the southern part of our FOV are higher than
those in the northern part of our FOV.

The ionization states for the ejecta component in P15 and P16 also 
reached the CIE condition.  Apart from P15, P16, and P17, the ionization
states for both components are in NEI condition in almost all of the cells.  
We found that all of the elements were distributed asymmetric to the
geometric center.  The EMs for O, Ne, and Mg are enhanced in P8 while
no enhancements are seen for the other elements there.  Also, the EMs
for all the elements other than Mg show enhancements in P15 and P16.

We obtained EMs of O, Ne, Mg, Si, S, and Fe for the ejecta component in
119 cells.  Multiplying the EM by the area of each cell, we obtained the
emission integral (hereafter EI, EI$=\int n_\mathrm{e}n_\mathrm{H} dV$,
$dV$ is the X-ray-emitting volume) of all elements for each cell,
and summed up the EIs for all cells within each FOV.
The summed-up EIs are summarized in table~\ref{tab:EI}.  

\begin{table*}
%\tabletypesize{\tiny}
  \caption{Summary of EI ($=\int n_\mathrm{e} n_\mathrm{i} dV$ for the
 ejecta component in units of 10$^{52}$\,cm$^{-3}$) in each
 FOV.$^{\ast}$}\label{tab:EI}   
  \begin{center}
    \begin{tabular}{lccccccc}
\hline
Elements & P8 & P12 & P13 & P14 & P15 & P16 & P17 \\
\hline
O&  5.3$^{+0.8}_{-0.7}$ & 2.4$\pm$0.4 & 2.6$\pm$0.4&
     2.0$^{+2.3}_{-1.0}$& 2.9$\pm$0.6& 3.0$\pm$0.5 & 
     1.8$\pm$0.2\\ 
Ne& 1.2$\pm$0.1& 0.50$^{+0.09}_{-0.08}$& 0.26$\pm$0.08&
     0.26$\pm$0.12& 0.44$\pm$0.09& 
     0.51$\pm$0.08 & 0.30$\pm$0.05\\ 
Mg& 0.26$\pm$0.05& 0.12$\pm$0.02& 0.10$\pm$0.03&
     0.06$^{+0.05}_{-0.03}$& 0.036$^{+0.038}_{-0.025}$& 
     0.03$\pm$0.02& 0.06$\pm$0.03\\
Si& 0.25$\pm$0.10& 0.24$\pm$0.03& 0.49$\pm$0.06&
     0.7$\pm$0.1& 1.3$\pm$0.2& 
     1.2$\pm$0.1& 0.5$\pm$0.2 \\
S&  0.18$\pm$0.07& 0.23$\pm$0.03& 0.40$\pm$0.04& 0.63$\pm$0.06&
     0.9$\pm$0.1& 
     0.9$\pm$0.1& 0.18$^{+0.10}_{-0.09}$\\ 
Fe& 0.24$\pm$0.04& 0.29$\pm$0.02& 0.44$\pm$0.01&
     0.57$\pm$0.02& 0.89$\pm$0.02& 
     0.82$\pm$0.02& 0.30$\pm$0.01\\ 
\hline
& &  \\[-8pt]
  \multicolumn{3}{@{}l@{}}{\hbox to 0pt{\parbox{140mm}{\footnotesize
\par\noindent
\footnotemark[$*$]Errors quoted are the mean 90\% confidence level for
     each cell.
\par\noindent
}\hss}}

    \end{tabular}
  \end{center}
\end{table*}

%Assuming that the ejecta component is confined in a spherical region of
%R $<$ 75$'$, i.e., 90\% of the radius of the Cygnus Loop, we estimated the
%plasma depth for each cell.  Then, we calculated the electron density
%based on the assumption that $n_\mathrm{e} = n_\mathrm{H}$.  Using the
%elemental abundances obtained, we calculated the masses of O, Ne, Mg, Si,
%S, and Fe for all the cells.  We summed up the mass from all the cells
%in each FOV and showed the summary in table~\ref{tab:mass}.  

\section{Discussion and Conclusions}

We observed the Cygnus Loop from NE to SW with Suzaku in seven
pointings.  Dividing the entire FOV into 119 cells, we extracted
spectra from all of the cells, and performed spectral analyses for them. 
Almost all of the spectra were significantly better fitted by a
two-component NEI model, rather than a one-component NEI
model. Judging from the abundances, the high-$kT_\mathrm{e}$ component
must be ejecta, while the low-$kT_\mathrm{e}$ component comes from the
swept-up matter.   

\subsection{Swept-up-Matter Distribution}

The temperature for the swept-up matter component is significantly
lower than that for the ejecta component.  On the other hand, it is
similar to that obtained for the rim of the Loop, where we expect no
contamination of the ejecta (e.g., Miyata et al.\ 2007). 
Therefore, we believe that we surely separated the X-ray emission of the
ejecta inside the Loop from that of the surrounding matter.  The EM
distribution of the swept-up matter (EM$_\mathrm{L}$) is inhomogeneous
in our FOV, as shown in figure~\ref{fig:param}.  The shell of the swept-up
matter seems to be thin in the SW part 
(i.e., P15 and P16) of the Loop relative to that in the NE part.  Such
a trend is also reported by XMM-Newton observations covering just north
of our Suzaku path (Tsunemi et al.\ 2007).  As Tsunemi 
et al.\ (2007) mentioned, there might be a blowout in the direction of
our line of sight around P15 and P16, such as a south blowout of the Loop
(see, figure~\ref{fig:hri_image}). 
The relatively high temperature in P15 and P16 suggests that the
velocity of the blast wave is higher than that in the other regions.
This fact indicates that the density of ambient matter in P15 and P16 is
lower than that in the other regions, supporting the thin
swept-up-matter shell.  

\subsection{Metal Distribution in Ejecta Component}

We divided our FOV into two parts: the NE part (P8, P12, P13, and P14) 
and the SW part (P15, P16, and P17).  There is a gap between P16 and
P17, which we did not observe.  Assuming EIs of the ejecta in the gap
to be averages of those in P16 and P17, we calculated the ratios
between EIs in the NE part and those in the SW part.  They are
O$\sim$1.2, Ne$\sim$1.3, Mg$\sim$3.2, 
Si$\sim$0.44, S$\sim$0.57, and Fe$\sim$0.60.  For simplicity, we assumed
that the mass ratios of those elements are equal to the EI ratios.
Regarding the O-Ne group, similar amounts of those
elements exist in the NE and SW parts.  In contrast to the O-Ne group, the
other elements show non-uniform distributions: Mg is distributed more in the
NE part by a factor of $\sim$3, while Si, S, and Fe are distributed more in the
SW part by a factor of $\sim$2. (Note that the mass ratio
being proportional to a product of the density times the emitting
volume will be lower than that of the EI ratio.)

A natural explanation for the asymmetries is an asymmetry at the time of
the SN explosion of the Cygnus Loop.   The degree of north-south (1:2)
asymmetry of the innermost ejecta, such as Si-, S-, and Fe-ejecta, is
consistent with that expected from asymmetric explosions resulting from
hydrodynamic instabilities, which are described in recent theoretical
models of SN explosions (e.g., Burrows et al.\ 2007).  
However, we should keep in mind that there are some other possibilities
that can produce the asymmetries of the ejecta.  The X-ray emission is
sensitive to both temperature and density, i.e., dense ejecta emit more
strongly than do thin ejecta and cooled ejecta will not continue to emit
X-rays.  Furthermore, an asymmetric environment that might be created by
stellar winds (e.g., Blondin et al.\ 1996; Dwarkadas 2007) as well
as hydrodynamic instabilities during the remnant evolution (
Jun \& Norman 1995) can make the ejecta distribution distorted.

If the mass ejection were to be anisotropic, it would lead a recoil of
a stellar remnant due to momentum conservation. 
Therefore, we expect to see the velocity of the stellar remnant directed
opposite to the momentum of the gaseous SN ejecta (e.g.,
Scheck et al.\ 2006).  In fact, recent observations of a stellar remnant
in Puppis~A SNR revealed that its velocity was very high
($>$1000\,km\,sec$^{-1}$) and directed opposite to the momentum of
fast-moving O-rich ejecta (Winkler \& Petre 2006; Hui \& Becker 2006).  
Since the Cygnus Loop is believed to be a remnant of a core-collapse SN
explosion, the presence of a stellar remnant is strongly expected.
So far, we have not found any stellar remnant associated with the SN
explosion of the Cygnus Loop (e.g., Miyata et al.\ 2001).  Taking into
consideration that the momentum of innermost ejecta (which is considered
to be strongly related to the stellar remnant) seems to be directed
toward south of the Loop, we might find the stellar remnant
associated with the Cygnus Loop in the north of the Loop.
We observed the Cygnus Loop from the NE rim to the SW rim with Suzaku
as well as XMM-Newton. The observation paths covered about one seventh of
the entire Loop.  Further, Suzaku and/or XMM-Newton observations of the
rest of the area are required to obtain the relative abundances for
the total ejecta as well as to reveal the ejecta structure in the
entire Loop.

\bigskip

This work is partly supported by a Grant-in-Aid for Scientific
Research by the Ministry of Education, Culture, Sports, Science and
Technology (16002004).  This study is also carried out as part of
the 21st Century COE Program, \rq{Towards a new basic science:
depth and synthesis}\rq.  S. K. is supported by a JSPS Research
Fellowship for Young Scientists.

%\newpage
%\newpage

%\newpage
%\newpage
%\newpage

%%%%%%%%%%%%%%%%%%%%%%%%%%%%%%%%%%%%%%%

%%%
% See the manual for the detail.
%%%

\end{document}